\documentclass{pasj00}
\draft

\begin{document}
\SetRunningHead{S. Katsuda et al.}{Abundance Inhomogeneity in the NE Rim of
the Cygnus Loop} 
\Received{2007/06/12}%{yyyy/mm/dd}
\Accepted{2007/08/07}%{yyyy/mm/dd}

\title{Abundance Inhomogeneity in the Northeastern Rim of
the Cygnus Loop Revealed by Suzaku Observatory}  

%%% begin:list of authors
% Do NOT capitalize all letters in "textsc".
\author{Satoru \textsc{Katsuda}, Hiroshi \textsc{Tsunemi}, Hiroyuki
\textsc{Uchida}, Emi \textsc{Miyata}, and Norbert \textsc{Nemes}} %
%  \thanks{Example: Present Address is xxxxxxxxxx}}
\affil{Department of Earth and Space Science, Graduate School of
  Science, Osaka University, 1-1 Machikaneyama, Toyonaka, Osaka
  560-0043
  }\email{katsuda@ess.sci.osaka-u.ac.jp,tsunemi@ess.sci.osaka-u.ac.jp,uchida@ess.sci.osaka-u.ac.jp,miyata@ess.sci.osaka-u.ac.jp,nnemes@ess.sci.osaka-u.ac.jp}

\author{Eric D. \textsc{Miller}}
\affil{Kavli Institute for Astrophysics and Space Research, Massachusetts
Institute of Technology, Cambridge, MA 02139,
U.S.A.}\email{milleric@space.mit.edu}  

\author{Koji \textsc{Mori}}
\affil{Department of Applied Physics, Faculty of Engineering,
University of Miyazaki, 889-2192, Japan}\email{mori@astro.miyazaki-u.ac.jp}

\and
\author{John. P. {\sc Hughes}}
\affil{Department of Physics and Astronomy, Rutgers University, 136
Frelinghhuysen Road, Piscataway, NJ 08854-8019,
U.S.A.}\email{jackph@physics.rutgers.edu} 

%\author{Motohide {\sc Kokubun}}
%\affil{Institute of Space and Astronautical Science, Japan Aerospace
%Exploration Agency, 3-1-1 Yoshinodai, Sagamihara, Kanagawa, 229-8510, Japan }\email{kokubun@astro.isas.jaxa.jp}

%\and
%\author{F. Scott. {\sc Porter}}
%\affil{NASA Goddard Space Flight Center, Greenbelt, MD 20771,
%U.S.A.}\email{Frederick.S.Porter@gsfc.nasa.gov} 

%%% end:list of authors

%%% Please use the following style in case that sorting by 
%%% affilation is impossible. 
%
% \author{%
%   D-Firstname \textsc{D-Familyname}\altaffilmark{1}
%   E-Firstname \textsc{E-Familyname}\altaffilmark{1,2}
%   and
%   F-Firstname \textsc{F-Familyname}\altaffilmark{2}}
% \altaffiltext{1}{Address of Institute}
% \email{ddddd@xxx.xxx.xx.xx}
% \email{eeeee@xxx.xxx.xx.xx}
% \altaffiltext{2}{Address of Institute}

%% `\KeyWords{}' always has to be placed before `\maketitle'.
 %Do NOT move this preamble from here!

\maketitle

\begin{abstract}
We present the results of a spatially resolved spectral analysis from
four Suzaku observations covering the northeastern rim of the Cygnus
Loop. A two-$kT_\mathrm{e}$ non-ionization equilibrium (NEI) model fairly
well represents our data, which confirms the NEI condition of the
plasma there. The metal abundances are depleted relative
to the solar values almost everywhere in our field of view. We find
abundance inhomogeneities across the field: the northernmost region
(Region~A) has enhanced absolute abundances compared with other regions. 
In addition, the relative abundances of Mg/O and Fe/O in Region~A are lower 
than the solar values, while those in the other regions are twice higher
than the solar values.  As far as we are concerned, neither a circumstellar
medium, (nor) fragments of ejecta, nor abundance inhomogeneities of the
local interstellar medium around the Cygnus Loop can explain the
relatively enhanced abundance in Region~A.  This point is left as an
open question for future work.
\end{abstract}

\section{Introduction}

The Cygnus Loop is a nearby (540\,pc: Blair et al.\ 2005) proto-typical
middle-aged ($\sim$10000 yr) supernova remnant (SNR) located at ($l,
b$)=(74$^\circ$, $-$\timeform{8.5D}). 
The foreground neutral hydrogen column density, $N_\mathrm{H}$, is
estimated to be $\sim0.04\times10^{22} \mathrm{cm}^{-2}$
(Inoue et al.\ 1980; Kahn et al.\ 1980).  The low foreground absorbing
material as well as the large apparent size (\timeform{2.5D} 
$\times$ \timeform{3.5D}: Levenson et al.\ 1997; Aschenbach \& Leahy
1999) and high surface brightness enable us to study the soft X-ray
emission from the Cygnus Loop.   

Miyata et al.\ (1994) observed the northeastern (NE) rim of the Cygnus Loop 
with ASCA.  They found non-equilibrium ionization (NEI)
conditions and depleted metal abundances relative to the solar
values.  The low metal abundances led them to consider that the plasma
in the NE-rim of the Cygnus Loop originated from swept-up matter,
rather than SN ejecta.  Recently, Miyata et al.\ 
(2007) (hereafter, Paper {\scshape I}) observed the same region with Suzaku
(Mitsuda et al.\ 2007) and performed spectral analysis from 2$'$ thick annular
regions.  They confirmed the metal deficiency as
well as the NEI conditions there.  Furthermore, the extended 0.2--12\,keV
energy range of the Suzaku X-ray CCD XIS camera (Koyama et al.\ 2007),
combined with its superior energy resolution, allowed them to detect
emission lines from highly ionized C and N for the first time from the
Cygnus Loop.   

\begin{figure*}
  \begin{center}
    \FigureFile(80mm,80mm){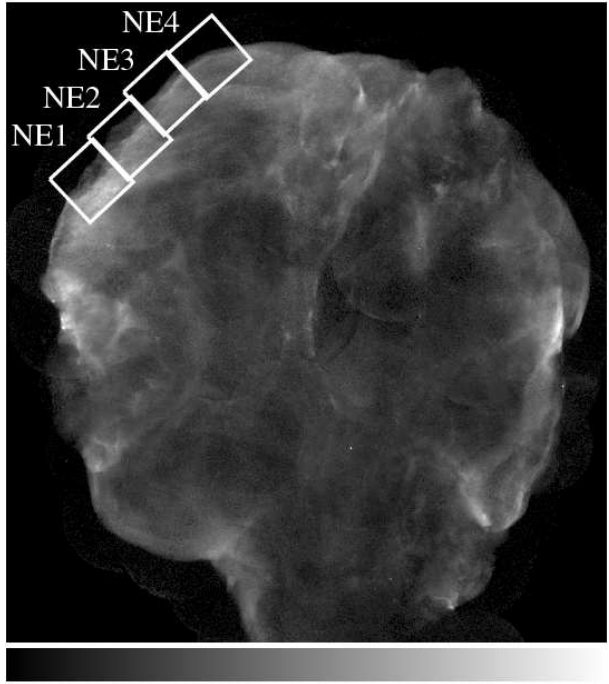}
    %%% \FigureFile(width,height){filename}
  \end{center}
  \caption{ROSAT HRI image of the entire Cygnus Loop. The Suzaku FOV
  from NE1 to NE4 are shown as white rectangles. }\label{fig:hri_image}
\end{figure*}

Using the Suzaku satellite, we observed the NE-rim of
the Cygnus Loop in four pointings (NE1--4) during the science
working group observing time.  The fields of view (FOV) are shown in 
figure~\ref{fig:hri_image}.  In Paper I, we presented the results
of an analysis for the NE2~region.  We here present the results of the
analyses for all four FOV using improved response files.

%\newpage

\section{Observations and Data Screening}

\begin{figure*}
  \begin{center}
    \FigureFile(80mm,80mm){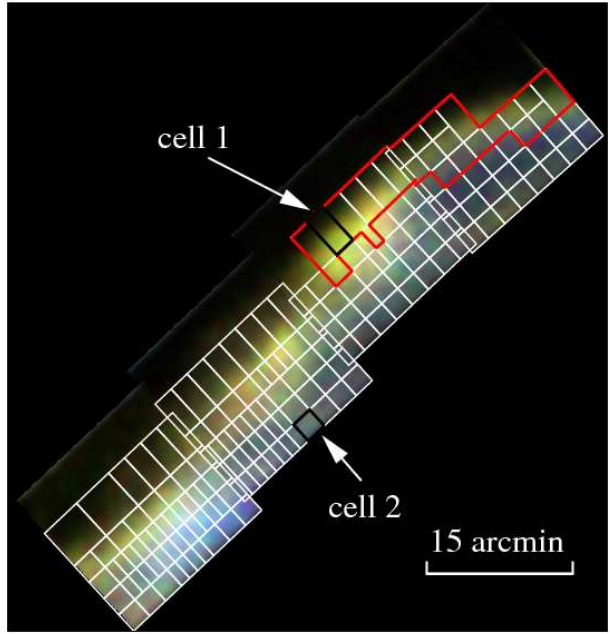}
    %%% \FigureFile(width,height){filename}
  \end{center}
  \caption{Merged Suzaku XIS1 three-color image of the four pointings
 (Red: 0.31--0.38\,keV band, i.e., C {\scshape VI} K$\alpha$, Green:
 0.38--0.46\,keV band, i.e., N {\scshape VI} K$\alpha$,
 Blue: 0.60--0.69\,keV band, i.e., O {\scshape VIII} K$\alpha$).  The
 data were binned by 8 pixels and smoothed by Gaussian
 distribution of $\sigma = 25^{\prime\prime}$. The 
 effects of exposure, vignetting, and contamination were corrected.  A
 small white rectangles are the cells where we extracted spectra.  We show
 example spectra from cells~1 and 2 in figure~\ref{fig:ex_spec}.  The red
 polygon identifies Region~A (see text).}
	\label{fig:xis_image} 
\end{figure*}

The observations were performed on 2005 November 23, 24, 29, and 30 for
NE1 (Obs. ID 500020010), NE2 (Obs. ID 500021010), NE3 (Obs. ID
500022010), and NE4 (Obs. ID 500023010), respectively.  We employed
revision 1.2 of the cleaned event data, and excluded the time region
where the attitude was unstable. Furthermore, we excluded data taken in
the low cut-off rigidity $<$\,6\,GV.  The net exposure time was 84\,ks
for all four observations after screening. We subtracted a blank-sky
spectrum obtained from the Lockman Hole, since the observation date of
the Lockman Hole (2005 November 14; Obs ID 100046010) was close to that
of the Cygnus Loop. 
Figure~\ref{fig:xis_image} shows the Suzaku XIS1 (back-illuminated CCD;
BI CCD) three-color image. For spectrum fitting, we used photons in the
energy range of 0.2--3.0\,keV for XIS1 and 0.4--3.0\,keV for XIS0, 2, and 3 
(front-illuminated CCD; FI CCD).  

\section{Spatially Resolved Spectral Analysis}

We divided the entire FOV into 184 cells (shown as small rectangles in
Figure~\ref{fig:xis_image}), such that each cell contains
2500--5000 photons for XIS0 to equalize the statistics.
We extracted spectra from them and performed spectral
analysis.  We can investigate the plasma structures along the azimuthal
direction as well as the radial direction from this analysis. 
Since the energy scale was not perfectly calibrated, we manually
adjusted the energy scale by shifting the energy within the uncertainty
of the calibration ($\pm5$\,eV; Koyama et al.\ 2007), so that
we could obtain better fits.  In order to generate a response matrix
file (RMF) and an ancillary response file (ARF), we employed {\tt
xisrmfgen} (Ishisaki et al.\ 2007) and {\tt xissimarfgen} 
(version 2006-10-26), respectively.
The low-energy efficiency of the XIS's shows degradation caused by 
contaminants accumulated on the optical blocking filter
(Koyama et al.\ 2007). This was taken into account when generating 
the ARF file. 

\begin{figure*}
  \begin{center}
    \FigureFile(80mm,80mm){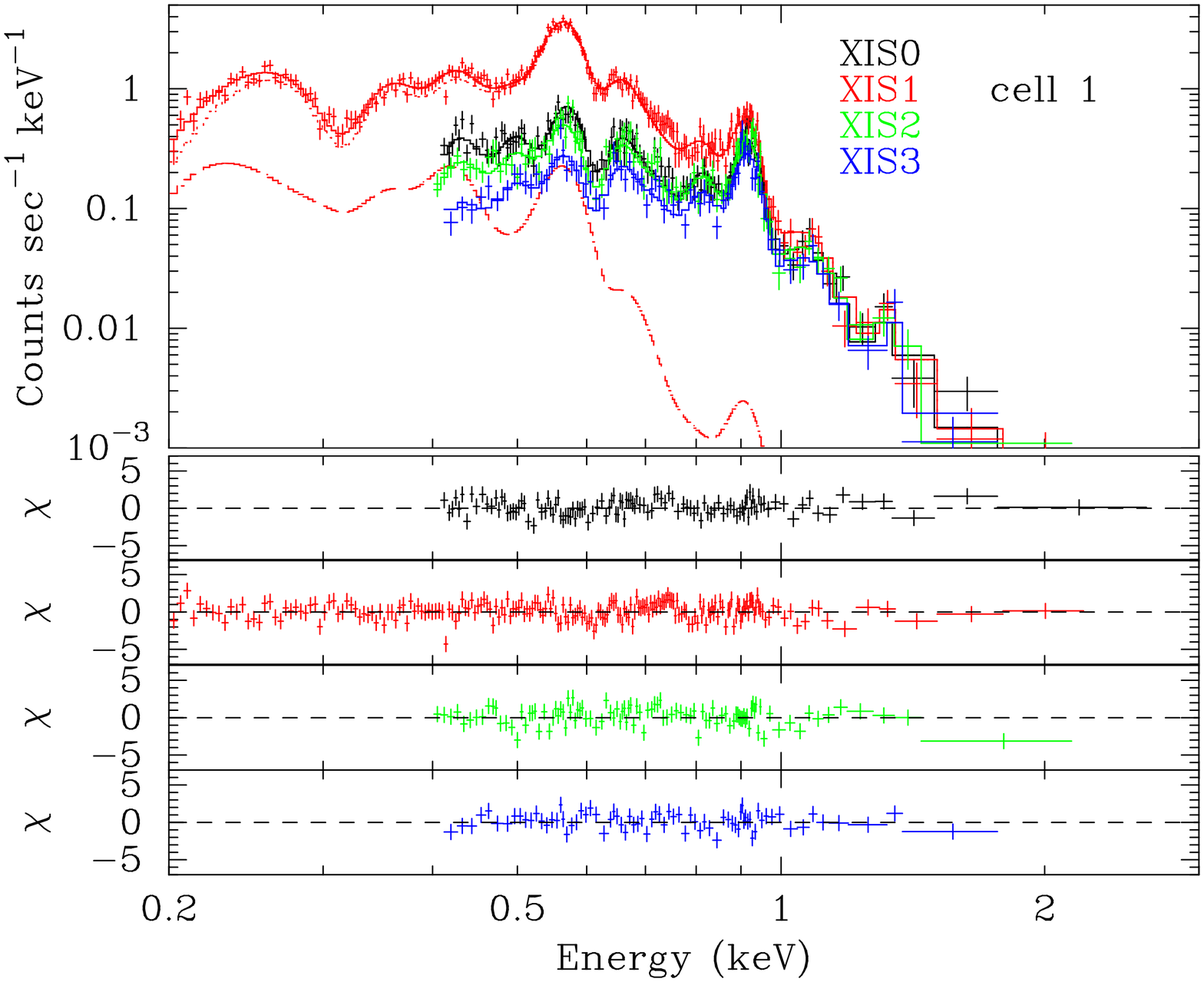}
    \FigureFile(80mm,80mm){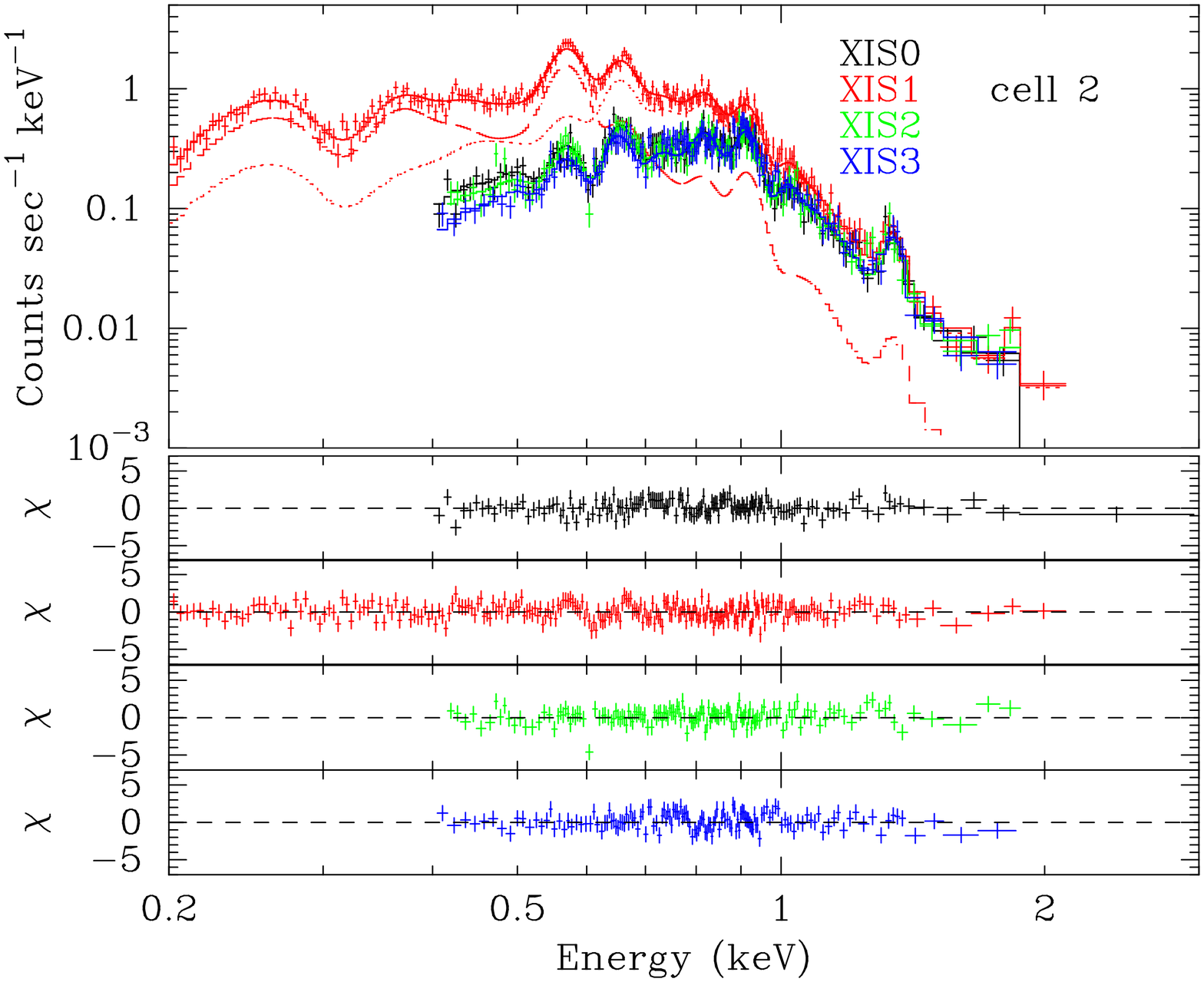}
    %%% \FigureFile(width,height){filename}
  \end{center}
  \caption{Left: X-ray spectra extracted from cell~1 in
 figure~\ref{fig:xis_image}.  The best-fit curves are shown with solid
 lines for the four XIS's.  The contribution of each component is
 shown by dotted lines only for XIS1.  The dashed line represents
 the low-temperature component while the dotted line represents the
 high-temperature component.  The lower panels show the residuals.
 Right: Same as left, but for cell~2.}\label{fig:ex_spec}    
\end{figure*}

\begin{figure*}
  \begin{center}
    \FigureFile(160mm,100mm){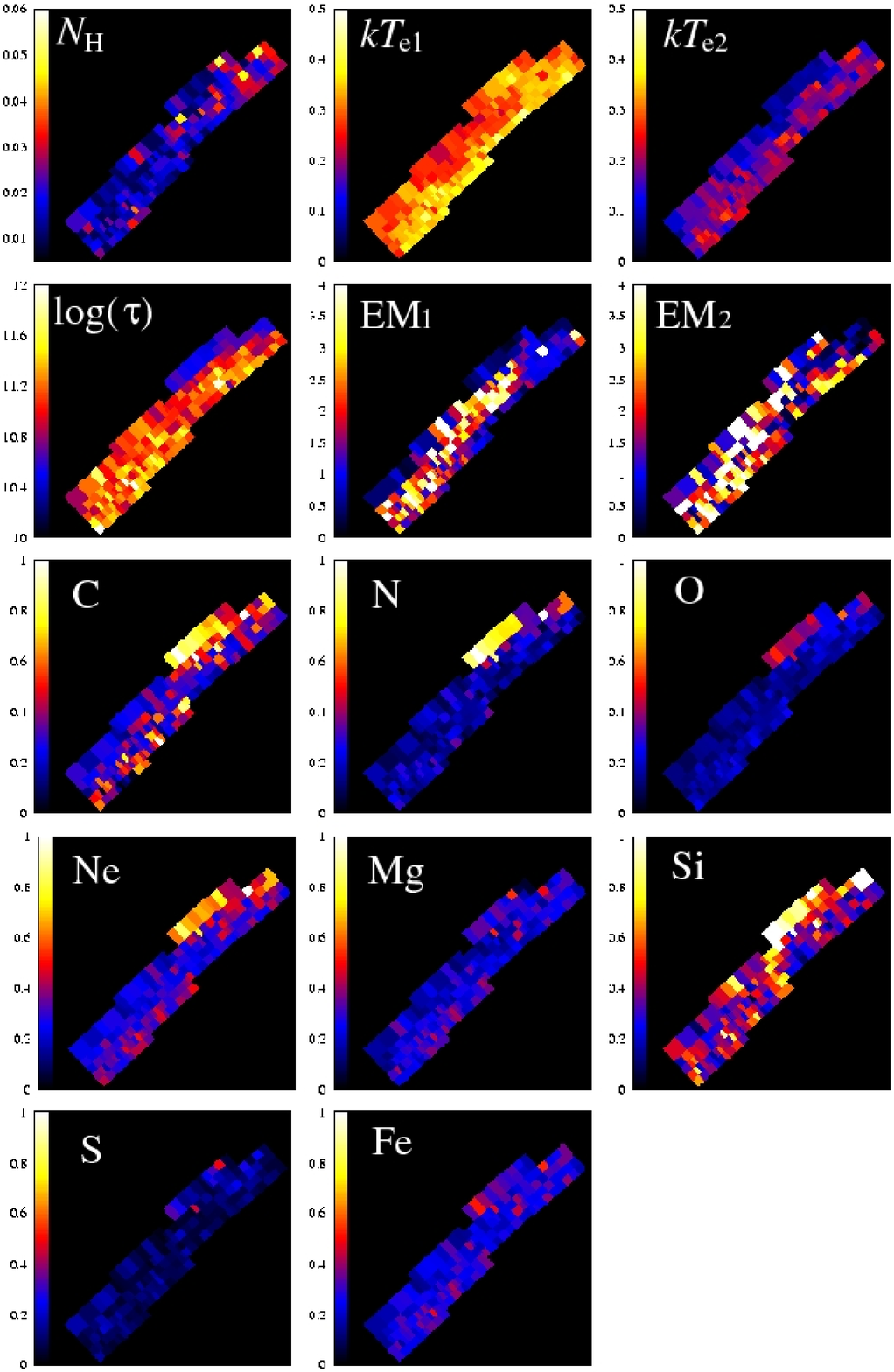}
    %%% \FigureFile(width,height){filename}
  \end{center}
  \caption{Maps of the best-fit parameters.  The units are $10^{22}
 \mathrm{cm}^{-2}$ for $N_\mathrm{H}$, keV for $kT_\mathrm{e1}$ and 
$kT_\mathrm{e2}$, cm$^{-3}$\,s for $\tau$, 
$10^{19} \mathrm{cm}^{-5}$ for EM$_\mathrm{1}$ and 
EM$_\mathrm{2}$, and solar values for abundances.}
\label{fig:param}   
\end{figure*}

Since the analysis in Paper-{\scshape I} already revealed that at least
two NEI components with different $kT_\mathrm{e}$ were required to
represent the spectra, we applied an absorbed
two-$kT_\mathrm{e}$-component NEI model for all spectra (the wabs
(Morrison \& McCammon 1983) and VNEI model (NEI version 2.0) in XSPEC
v\,11.3.1; e.g., Borkowski et al.\ 2001).  The free 
parameters were $N_\mathrm{H}$; electron temperature, $kT_\mathrm{e}$;
ionization timescale, $\tau$; emission measure, EM (EM$=\int
n_\mathrm{e}n_\mathrm{H} dl$, where $n_\mathrm{e}$ and $n_\mathrm{H}$
are the number densities of electrons and protons, respectively and $l$ is
the plasma depth); abundances of C, N, O, Ne, Mg, Si, S, Fe, and
Ni. We set the abundance of Ni equal to that of Fe.  The other
elemental abundances were fixed to the solar values (Anders \&
Grevesse 1989). 
We individually varied $kT_\mathrm{e}$ and EM while other parameters were
tied in the two components.  We confined the variation of $N_\mathrm{H}$
to be 0.01 to 0.06$\times10^{22}\mathrm{cm}^{-2}$ (Inoue et al.\ 1980;
Miyata et al.\ 2007).  We here refer to this model as a VNEI1 model.
It gave us fairly good fits for all
spectra (reduced $\chi^2$ ranges from 0.90 to 1.27). 
The fit statistics are dramatically improved compared to those obtained in
Paper {\scshape I} (maximum reduced $\chi^2$ of 2.81).
This is mainly due to the fact that the post-launch degradation of the
XIS energy resolution is now included in our spectral-response function,
which was not possible at the time Paper {\scshape I} was written. 
Figure~\ref{fig:ex_spec} shows example spectra from cells~1 and 2 in
figure~\ref{fig:xis_image} with the best-fit models.  The best-fit
parameters for the cells are summarized in table~\ref{tab:ex_param} (VNEI1).
Maps of the best-fit values are presented in figure~\ref{fig:param}.

\begin{table*}
%\tabletypesize{\tiny}
  \begin{center}
  \caption{Spectral-fit parameters for cells~1 and 2.}\label{tab:ex_param}
    \begin{tabular}{lccccc}
      \hline
Parameter & cell~1 (VNEI1) & cell~1 (VNEI2) & cell~1 (VPSHOCK)& cell~2
     (VNEI1) & cell~2 (VNEI2)\\ 
\hline
$N_\mathrm{H}$[$\times10^{22}$cm$^{-2}$]\dotfill &
     0.024$\pm0.001$&0.023$^{+0.02}_{-0.01}$&0.024$^{+0.003}_{-0.001}$&
     $<0.021$ &$<0.020$ \\
$kT_\mathrm{e1}$[keV] \dotfill & 0.27$\pm$0.01 & 0.27$\pm$0.01&0.26$\pm$0.01&0.39$\pm0.02$&0.39$\pm$0.01\\
$kT_\mathrm{e2}$[keV] \dotfill & 0.09$\pm$0.01 & 0.08$^{+0.02}_{-0.04}$& 0.08$\pm$0.01&0.21$\pm0.02$&0.23$\pm$0.02\\
C \dotfill& 1.06$\pm$0.08 &1.03$\pm$0.08&1.2$\pm$0.1&0.24$\pm$0.04&0.20$\pm$0.04\\
N \dotfill& 1.03$\pm$0.06&1.04$\pm$0.06&1.08$^{+0.1}_{-0.04}$&0.09$\pm$0.02&0.10$\pm$0.02\\
O \dotfill& 0.53$\pm$0.02&0.53$\pm$0.02&0.54$\pm$0.02&0.131$\pm$0.004&0.131$\pm$0.004\\
Ne \dotfill& 0.84$\pm$0.04&0.84$\pm$0.04&0.91$^{+0.03}_{-0.05}$ &0.29$\pm$0.01&0.28$\pm$0.01\\
Mg \dotfill& 0.35$\pm$0.12& 0.35$\pm$0.12&0.39$^{+0.13}_{-0.16}$ &0.20$\pm$0.03&0.20$\pm$0.03\\
Si \dotfill& 1.9$\pm$0.2& 1.8$\pm$0.2&1.9$^{+0.14}_{-0.15}$&0.24$\pm$0.04&0.24$\pm$0.04\\
S \dotfill& $<$0.5&$<$0.5&$<$0.4&0.14$\pm$0.10&0.16$\pm$0.12\\
Fe(=Ni) \dotfill&0.52$\pm$0.04&0.52$\pm$0.04&0.60$^{+0.06}_{-0.03}$ &0.22$\pm$0.01&0.21$\pm$0.01\\
log$(\tau /\mathrm{cm}^{-3}\,\mathrm{s})$\dotfill
     &10.65$\pm0.08$ &\dotfill&\dotfill&11.21$^{+0.09}_{-0.10}$&\dotfill\\
log$(\tau_1 /\mathrm{cm}^{-3}\,\mathrm{s})$\dotfill
     &\dotfill &10.63$\pm$0.04&\dotfill&\dotfill&11.24$^{+0.05}_{-0.06}$\\
log$(\tau_2 /\mathrm{cm}^{-3}\,\mathrm{s})$\dotfill
     &\dotfill&10.6$<$&\dotfill&\dotfill&11.0$\pm$0.1\\
log$(\tau_\mathrm{lower} /\mathrm{cm}^{-3}\,\mathrm{s})$\dotfill
     &\dotfill&\dotfill&0 (fixed)&\dotfill&\dotfill\\
log$(\tau_\mathrm{upper} /\mathrm{cm}^{-3}\,\mathrm{s})$\dotfill
     &\dotfill&\dotfill&11.00$\pm$0.03&\dotfill&\dotfill\\
EM$_1$[$\times10^{19}$ cm$^{-5}$]\dotfill& 0.48$\pm0.01$& 0.48$\pm0.01$&0.52$\pm$0.02 &1.4$\pm$0.04&1.39$\pm$0.04\\ 
EM$_2$[$\times10^{19}$ cm$^{-5}$]\dotfill& 0.9$\pm$0.2& 0.9$\pm$0.2
     &0.5$\pm$0.2&2.4$\pm$0.1 &2.4$\pm$0.1\\  
\hline
$\chi^2$/d.o.f. \dotfill & 611/508& 611/507 & 619/508 &706/699 &706/698\\
      \hline
\\[-8pt]
  \multicolumn{3}{@{}l@{}}{\hbox to 0pt{\parbox{140mm}{\footnotesize
     \par\noindent 
\footnotemark[$*$]Other elements are fixed to those of solar values.\\
     The values of abundances are multiples of solar value.\\  The errors
     are in the range $\Delta\,\chi^2\,<\,2.7$ on one parameter.  \\ 
     The subscript 1 denotes the high temperature component \\
     while 2 denotes the low temperature component.
\par\noindent 
%\footnotemark[$\dagger$]EM denotes the emission measure $\int
%     n_\mathrm{e} n_\mathrm{H} dl$. 
}\hss}}

    \end{tabular}
  \end{center}
\end{table*}

\section{Results}

\subsection{Abundances}

The abundances of O, Ne, Mg, and Fe are consistent with those in
Paper-{\scshape I}. On the other hand, C and N in our analysis are
systematically higher, while Si and S are systematically lower than those
in Paper {\scshape I}.  There are two main reasons that can explain
this discrepancy. Firstly, we allowed the abundance of S to vary freely in
our models, while in Paper {\scshape I} it was fixed to the solar value.
Since strong emission lines of S L fall around the C K 
($\sim$0.35\,keV) and N K ($\sim$0.5\,keV) emission lines, the 
abundances of C and N are affected by that of S.  In our 
spectral analysis, the typical abundance of S is $\sim$0.2-times the 
solar value, resulting in higher abundances 
of C and N than those in Paper {\scshape I}.  Secondly, we 
included data in the Si K band (1.7--1.9\,keV), while in Paper {\scshape
I} excluded it due to a calibration uncertainty.  
Since the Si K line was not so strong in our data, we found that the
calibration uncertainty did not play an important role in our results. 
Due to limited available atomic data for these transitions, the
emissivities of the Si L lines in our models are expected to contain large
uncertainty relative to those of the Si K lines.  Therefore, we believe that
our results are more reliable than those in Paper {\scshape I} in
which the abundance of Si was determined by the emission lines of Si L. 

\begin{figure*}
  \begin{center}
    \FigureFile(160mm,100mm){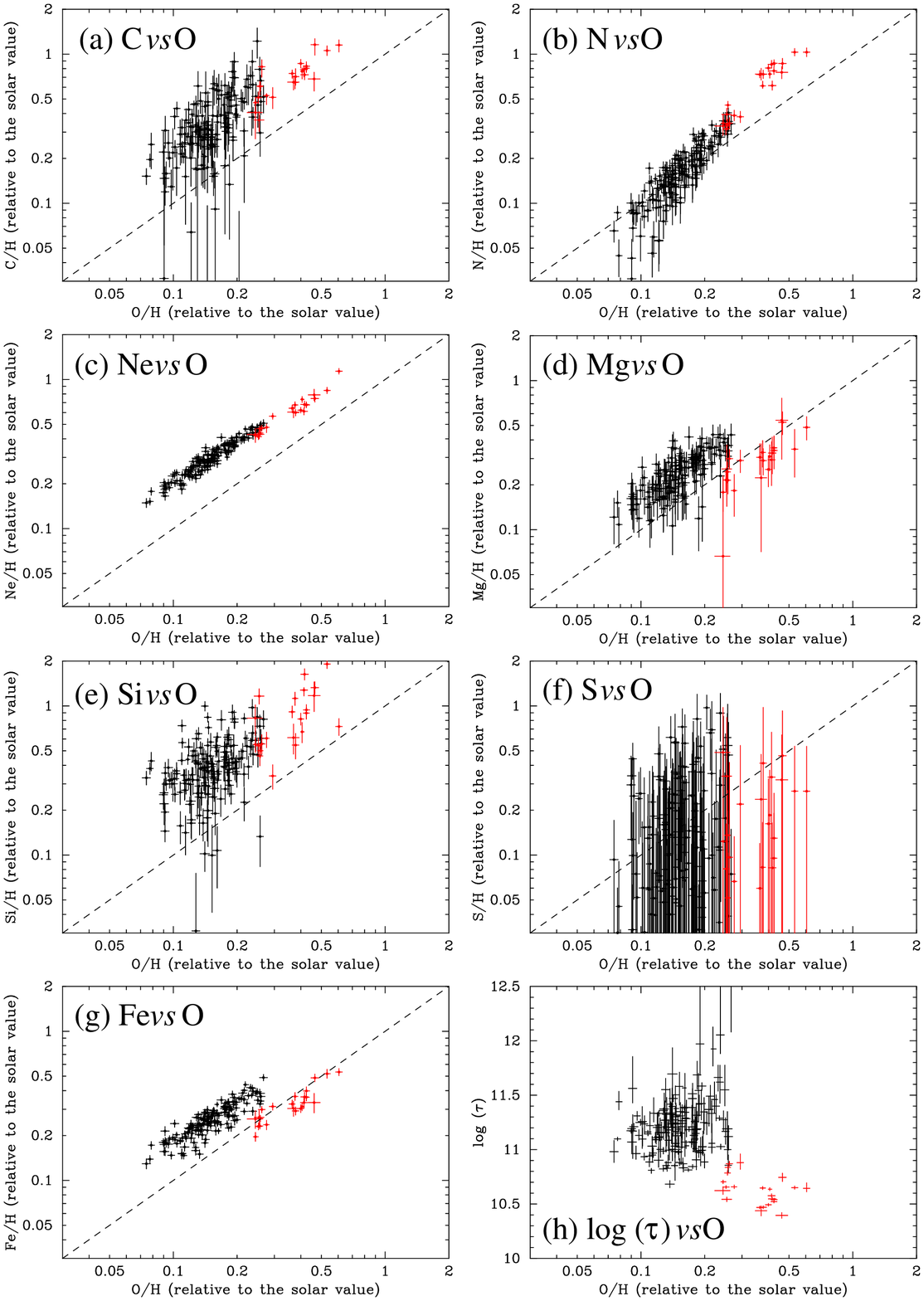}
    %%% \FigureFile(width,height){filename}
  \end{center}
  \caption{Each elemental abundance or ionization timescale versus O is
 plotted for the 184 cells. Red data points come from Region~A, while
 black points come from the remaining cells.  The dashed line from the
 upper right to the lower left represents the solar
 ratio.}\label{fig:correlation}     
\end{figure*}

We obtained abundances within 184 spatial cells, as shown in
figure~\ref{fig:param}. In Figures~\ref{fig:correlation}
(a)--\ref{fig:correlation}(g), we show plots of the correlations
between each elemental abundance versus the O abundance for all of these
regions. The absolute abundances (relative to H) for all the elements
vary significantly from cell to cell.  The relative abundances of
C, Ne, Si, and S to O are similar in all the cells, while
those of Mg and Fe are divided into two groups; one shows lower values
than the solar value (red crosses in figure~\ref{fig:correlation}), while the
other shows about two-times higher values than the solar value (black
crosses in figure~\ref{fig:correlation}).  We indicate the regions of red
crosses in a red polygon in figure~\ref{fig:xis_image}.  We found that those
regions were concentrated north of our FOV, where the absolute
abundances are relatively higher than those in other regions.
Hereafter we refer to the region outlined by the red polygon in
figure~\ref{fig:xis_image} as Region A.  We summarize mean abundances,
mean 90\% errors, and standard deviations of the abundances for both
Region~A and the other regions in table~\ref{tab:abund}.

In our model, the ionization timescale is imposed to be the same
for the two components for simplicity of the model, just as employed in
Paper {\scshape I}.  In this paragraph, we consider whether or not
the introduction of separate ionization timescales 
for the two components changes the results.  We fitted the spectra
in cells~1 and 2 with a two-component VNEI model whose free parameters
were the same as those used in the VNEI1 model, but used separate ionization
timescales for the two components that we call the VNEI2
model.  The best-fit parameters and fit statistics are summarized in
table~\ref{tab:ex_param} (VNEI2). 
We found that all of the best-fit parameters were consistent with
those obtained with the VNEI1 model.  We should note that the 
results are quite similar to those from the VNEI1 model.
We checked the value of $\chi^2$ as a function of $\tau_2$ over a large
range of values (i.e., $2\times10^{10}$--$1\times10^{12}$ cm$^{-3}$\,s),
and confirmed that the results were not due to secondary local minima.
We thus conclude that the obtained abundances are not affected by 
introducing a separate ionization timescale for the two
components.

Figure~\ref{fig:param} clearly shows that Region~A is 
the location of a significantly different best-fit value for the
ionization timescale.  We plot the correlation between the ionization timescale
and the O abundance in figure~\ref{fig:correlation} (h).  It shows 
anti-correlation, which causes us some worry about the derived abundances,
since the NEI models that we employ are really just very 
simplistic approximations to the true physical conditions under which
these plasmas emit.  Figure~\ref{fig:param} also indicates that Region~A
has a rapidly changing ionization state throughout: from $\tau = 0$
at the shock front to $\tau =~\sim10^{11}\,\mathrm{cm^{-3}\,s}$.
Our models, however, assume a single ionization timescale (VNEI).
In this context, we fitted the spectrum in cell~1 (which is within
Region~A) with the VPSHOCK model (e.g., Borkowski et al.\ 2001) in
XSPEC, which assumes a constant temperature and a distribution of the
ionization timescale, in which we take all ionization timescales up to a
fitted maximum value, starting from zero.  We employed a
two-temperature VPSHOCK model 
with the same number of free parameters as in the case of the VNEI model.
The best-fit values and fit statistics are summarized in
table~\ref{tab:ex_param}.  We find that the abundances are almost
equal between the VNEI model and the VPSHOCK model, which supports the
idea that the abundances in Region~A are really different from those in
the rest of our FOV.

\begin{table*}
%\tabletypesize{\tiny}
  \caption{Mean elemental abundances in Region~A and the rest of the
 region.$^{\ast}$}\label{tab:abund} 
  \begin{center}
    \begin{tabular}{lcc}
      \hline
Parameters & Region~A & Excluding Region~A\\
\hline
C \dotfill&0.74$^{+0.06}_{-0.08}$ (0.26)& 0.36$^{+0.07}_{-0.08}$ (0.18)\\
N \dotfill&0.66$\pm$0.05 (0.27)& 0.17$\pm$0.02 (0.08)\\
O \dotfill&0.38$\pm$0.02 (0.13)& 0.16$\pm$0.01 (0.05)\\
Ne \dotfill&0.63$\pm$0.03 (0.22) & 0.31$\pm$0.02 (0.09)\\
Mg \dotfill&0.32$\pm$0.08 (0.12) & 0.24$\pm$0.04 (0.08)\\
Si \dotfill&0.9$\pm$0.1 (0.5) & 0.41 $\pm$0.06 (0.18)\\
S \dotfill& 0.2$\pm$0.2 (0.1)& 0.2$^{+0.2}_{-0.1}$ (0.2)\\
Fe(=Ni) \dotfill&0.34$\pm$0.02 (0.12) &0.26$\pm$0.01 (0.07)\\
\hline
& &  \\[-8pt]
  \multicolumn{3}{@{}l@{}}{\hbox to 0pt{\parbox{140mm}{\footnotesize
\par\noindent
\footnotemark[$*$]The values in brackets represent the standard
     deviations.\\  Errors quoted are mean values for each cell.
\par\noindent
}\hss}}

    \end{tabular}
  \end{center}
\end{table*}

\subsection{$kT_\mathrm{e}$, $\tau$, $N_\mathrm{H}$, and EM}

In the NE2~region, the values of both $kT_\mathrm{e1}$ and $kT_\mathrm{e2}$
increase from the outermost cells toward the innermost cells.
This trend is consistent with previous X-ray observations of the
relevant regions obtained with ASCA (Miyata et al.\ 1994;
Miyata \& Tsunemi 1999), XMM-Newton (Katsuda \& Tsunemi 2007; Tsunemi
et al.\ 2007), and Suzaku (Miyata et al.\ 2007).
The values are also quantitatively consistent with those in
Paper {\scshape I}.  Also, we confirmed that the ionization state  
is far from the CIE condition in the NE2~region, and found that  
the ionization state is in the NEI condition everywhere in the
NE1--4~regions.   
The ionization states in the outermost cells in the NE3 and 4~regions 
(which correspond to Region~A) are relatively lower than those in
the other regions.  The EMs for the hot component in the responsible
regions are relatively lower than those in the other regions, suggesting
low electron densities there.  Since almost all of the emission lines come
from the hot component (see, figure~\ref{fig:ex_spec} left), the
obtained ionization states represent those for the hot component.  
Therefore, the relatively low ionization states are likely due
to the low density.  
We found that the $N_\mathrm{H}$ for almost all regions is
around 0.02$\times10^{22}\mathrm{cm^{-2}}$, while we see about a
twice-enhanced column density in the NE4~region. 

\begin{figure*}
  \begin{center}
    \FigureFile(160mm,100mm){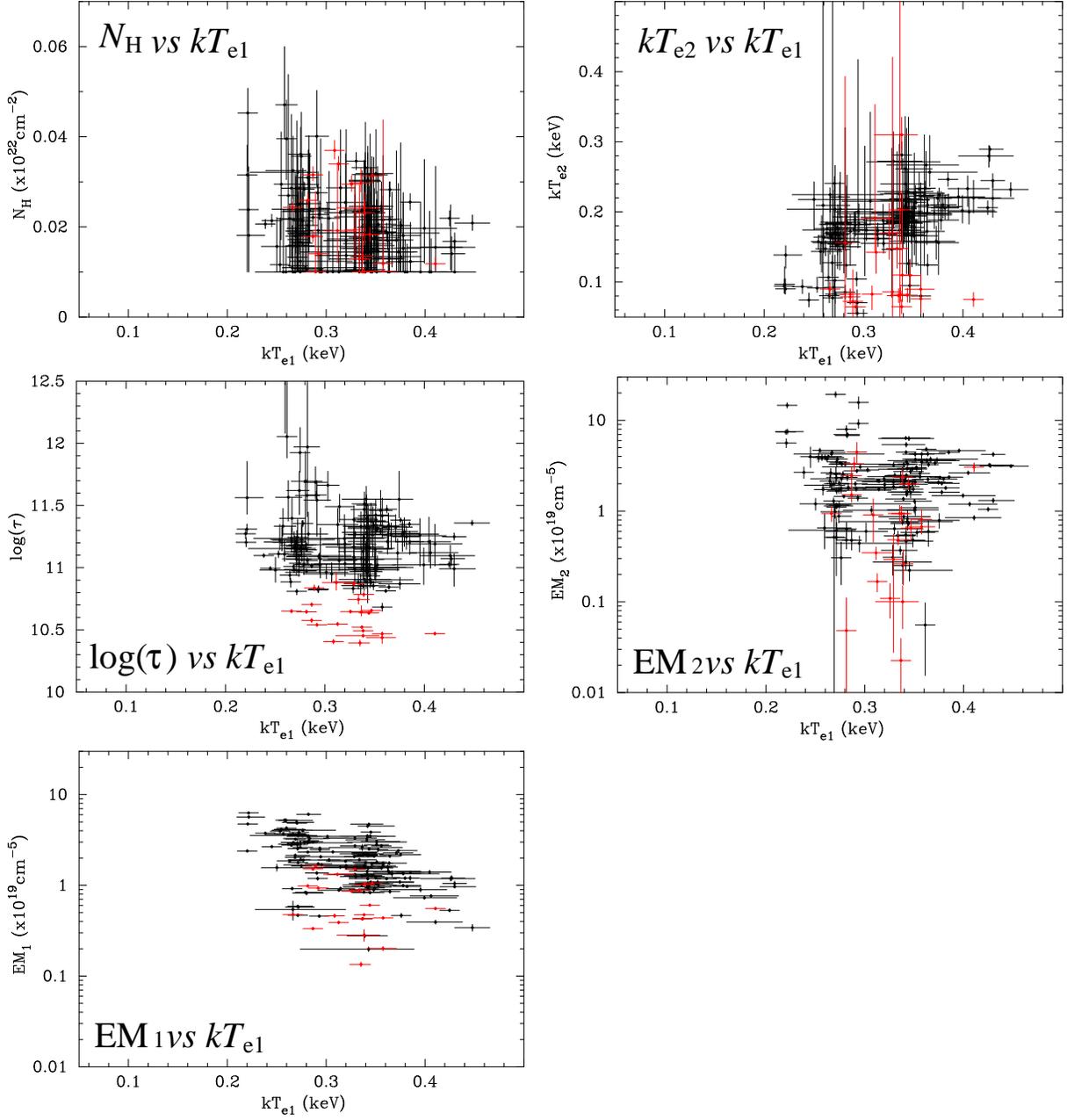}
    %%% \FigureFile(width,height){filename}
  \end{center}
  \caption{$N_\mathrm{H}$, $kT_\mathrm{e2}$, log($\tau$), EM$_2$, and
 EM$_1$ versus $kT_\mathrm{e1}$ are plotted for the 184 cells. Red data
 points come from Region~A while black points come from the remaining
 cells.}\label{fig:correlation2}    
\end{figure*}

To check the significance level of the observed variations for
$kT_\mathrm{e}$, $\tau$, $N_\mathrm{H}$, and EM, and also to look for
possible correlations among those parameters, we give plots of the
correlations between each parameter versus $kT_\mathrm{e1}$ in
figure~\ref{fig:correlation2}.  We found that all the parameters were
sufficiently constrained to confirm the observed variations in our FOV. We 
also found that the ionization states in Region~A are significantly
lower than those in the other region.  We could not find any
significant correlations among those parameters, although the
distribution of $kT_\mathrm{e1}$ is clustered around
$\sim$0.28\,keV and $\sim$0.35\,keV.  As shown in figure~\ref{fig:param},
the low- and high-temperature clusters are generally located in the
outer and inner regions of the remnant, respectively, which is at least 
qualitatively consistent with what we expect from Sedov-phase SNRs.

\section{Discussion and Conclusions}

We observed the northeastern rim of the Cygnus Loop with the Suzaku
observatory in four pointings. We then divided the FOV into 184 cells
and performed a spatially resolved spectral analysis.  Following the
analysis in Paper {\scshape I} in which the authors concluded that
there was a multi-temperature plasma along the line of sight, we
applied a two-component NEI model with different $kT_\mathrm{e}$
values for all of the spectra.   

Assuming a spherically-symmetric uniform emitting region, and that the
two temperature components are in pressure equilibrium, we can estimate
the plasma depth for the two components in each cell.  We calculated the
total masses of the low and high-temperature components in Region~A and the
rest of the region, respectively, to be $\sim$0.7\,M$_\odot$ and
$\sim$6\,M$_\odot$.  In these calculations, we assumed that
$n_\mathrm{e} = n_\mathrm{H}$ and the volume filling factor was unity. 

The relative abundances of C, Ne, Si, and S to O are almost constant in
our FOV, while those of Mg and Fe are divided into two groups: one 
shows lower values than the solar value, while the other shows
two-times higher values than the solar value.  
Regions with low relative abundances are concentrated into the very
northernmost cells (Region~A).  The absolute abundances in
Region~A turned out to be relatively higher than those in the rest of the
region.

The low abundances in regions other than Region~A confirm the
absence of SN ejecta contamination at the northeast rim, and
argues for a swept-up origin.  Depleted abundances at the rim
of the Cygnus Loop have now been reported by several X-ray
studies.  For example, an ASCA observation of the NE rim (which overlaps
the NE2~region here) revealed the abundance of O to be 0.2-times the
solar value (Miyata \& Tsunemi 1999).  A Chandra observation of the 
southwestern rim showed the O-group abundance to be 0.22-times the
solar value (Leahy 2004; in his spectral 
analysis, he fixed the abundances of C and N to be relatively the same
as O (O-group)).  Low abundances appear to be a common result of X-ray
spectral analysis of the rim of the Cygnus Loop, although an adequate
explanation for this result has yet to be proposed. 

The abundances in Region~A are higher than those in the rest of the
region by factors of $\sim$2.1 (C), $\sim$3.9 (N), $\sim$2.4 (O),
$\sim$2.1 (Ne), $\sim$1.3 (Mg), $\sim$2.4 (Si), $\sim$1.2 (S), and
$\sim$1.3 (Fe). Since there is evidence in many SNe that the
circumstellar medium (CSM) frequently shows enhanced abundance ratios of
N/C and N/O (Fransson et al.\ 2005; Chevalier 2005) relative to the
solar values as a result of CNO processing in progenitor stars, the
strongly enhanced abundance of N in Region~A  
relative to the rest of the region may lead us to consider that CSM
contamination is evident.  However, theoretical nucleosynthetic
calculations show that the CSM is only rich in N
abundance compared with the initial composition of a progenitor
star (e.g., \cite{Rauscher2002}). This cannot fully explain that all of
the metal abundances in Region~A are enhanced relative to those in the
rest of the region.  
Can fragments of ejecta from the SN explosion explain the enhanced
metal abundances?  Fragments of ejecta observed in many SNRs (e.g., Cas A:
Fesen et al.\ 2001, Vela: Aschenbach et al.\ 1995, Tycho:
Decourchelle et al.\ 2001) commonly show knotty shapes or head-tail
structures.  There is no such structure in Region~A, showing no
indication of fragments of ejecta.  We next consider the possibility
of abundance inhomogeneities of the local interstellar medium in the
vicinity of the Cygnus Loop.  Observations of an
optically thin 1356\,$\AA$ resonance line and H {\scshape I} Ly$\alpha$
in 13 sight lines showed remarkable homogeneity of the interstellar
gas-phase O/H ratio at a level of $\sim$5\% within
about 500\,pc of the Sun (Meyer et al.\ 1998).  More recent observations
for 36 sight lines confirmed the homogeneity of O/H abundance ratio and
revealed that the ratio is uniform within 800\,pc of the Sun
(Cartledge et al.\ 2004).  Therefore, it is difficult to explain such a strong
variation in the abundances at these very small scales by abundance
inhomogeneities of the local interstellar medium.  The nature of the
observed abundance inhomogeneity is left as an open question for future
work.

%It is well known that there is metallicity
%gradients in our galaxy. Assuming that the Cygnus Loop is 8.5\,kpc away
%from the galactic center, we calculated the metal abundances of N, O,
%and S to be 0.5$\sim$0.8, 0.3$\sim$0.6, and 0.3$\sim$1, respectively
%based on table~7 in Rudolph et al.\ (2006).  Then, we note that the
%abundances of N, O, and S in Region~A are consistent with those
%expected by metallicity gradients in our galaxy.   

\bigskip

This work is partly supported by a Grant-in-Aid for Scientific
Research by the Ministry of Education, Culture, Sports, Science and
Technology (16002004).  This study is also carried out as part of
the 21st Century COE Program, \lq{\it Towards a new basic science:
depth and synthesis}\rq.  S. K. is supported by a JSPS Research Fellowship
for Young Scientists.

%\newpage
%\newpage

%\newpage
%\newpage
%\newpage

%%%%%%%%%%%%%%%%%%%%%%%%%%%%%%%%%%%%%%%

%%%
% See the manual for the detail.
%%%


\begin{thebibliography}{}
% Journals(e.g. A\&A,ApJ,AJ,NMRAS,PASP ...)
% Authors, Year, Journal, Vol#, Page#
% Journal Title Abbreviation >> http://www.asj.or.jp/pasj/Jabb.html
%\bibitem[Aauthor et al.(2001)]{}
%   Aauthor, A., Bauthor, B., Cauthor, C.\ 2001, PASJ, vol, page
% Books
\bibitem[Anders \& Grevesse 1989]{Anders1989}
        Anders, E., \& Grevesse, N. 1989, Geochem. Cosmochim. Acta, 53, 197
\bibitem[Aschenbach et al.\ 1995]{Aschenbach1995}
        Aschenbach, B., Egger, R., \& Trumper, J. 1995, Nature, 373, 587
\bibitem[Aschenbach \& Leahy 1999]{Aschenbach1999}
	Aschenbach, B. \& Leahy, D. A. 1999, A\&A, 341, 602			 
%\bibitem[Blair et al.\ 1999]{Blair1999}
%        Blair, W. P., Sankrit, R., \& Raymond, J. C., and
%           Long. K. S. 1999, AJ, 118, 942
\bibitem[Blair et al.\ 2005]{Blair2005}
        Blair, W. P., Sankrit, R., \& Raymond, J. C. 2005, AJ, 129, 2268
%\bibitem[Borkowski et al.\ 1994]{Borkowski1994}
%        Borkowski, K. J., Sarazin, C. L., \& Blondin, J. M. 1994, ApJ,
%        429, 710
\bibitem[Borkowski et al.\ 2001]{Borkowski2001}
        Borkowski, K. J., Lyerly, W. J., \& Reynolds, S. P. 2001,
        ApJ, 548, 820
\bibitem[Cartledge et al.\ 2004]{Cartledge2004}
	Cartledge, S. I. B., Lauroesch, J. T., Meyer, D. M., \& Sofia,
				 U. J. 2004, ApJ, 613, 1037
\bibitem[Chevalier 2005]{Chevalier2005}
        Chevalier, R. A. 2005, ApJ, 619, 839
\bibitem[Decourchelle et al.\ 2001]{Decourchelle2001}
        Decourchelle, A. et al.\ 2001, A\&A, 365, L218
%\bibitem[Dopita et al.\ 1977]{Dopita1977}
%        Dopita, M. A., Mathewson, D. S., and Ford, V. L. 1977 ApJ, 214, 179
\bibitem[Fesen 2001]{Fesen2001}
	Fesen, R. A.\ 2001, ApJS, 133, 161
\bibitem[Fransson et al.\ 2005]{Fransson2005}
        Fransson, C. et al.\ 2005, ApJ, 622, 991
%\bibitem[Ghavamian et al.\ 2001]{Ghavamian2001}
%        Ghavamian, P., Raymond, John., Smith, R. C., \& Hartigan,
%        P. 2001, ApJ, 547, L995
%\bibitem[Hamilton et al.\ 1983]{Hamilton1983}
%        Hamilton, A. J. S.,  Chevalier, R. A., \& Sarazin, C. L. 1983,
%           ApJS, 51, 115
%\bibitem[Hayashida et al.\ 2007]{Hayashida2007}
%	Hayashida et al.\ 2007, this issue			 
%\bibitem[Hester et al.\ 1994]{Hester1994}
%        Hester, J. J., Raymond, J. C., \& Blair, W. P. 1994, ApJ, 420, 721
\bibitem[Ishisaki et al.\ 2007]{Ishisaki2007}
	Ishisaki, Y., et al.\ 2007, PASJ, 59, S113
\bibitem[Inoue et al.\ 1980]{Inoue1980}
        Inoue, H., Koyama, K., Matsuoka, M., Ohashi, T., Tanaka, Y.,
         Tsunemi, H. 1980 ApJ, 238, 886
\bibitem[Kahn et al.\ 1980]{Kahn1980}
        Kahn, S. M., Charles, P. A., Bowyer, S., \& Blissett, R. J. 1980,
        ApJ, 242, L19
\bibitem[Katsuda \& Tsunemi 2007]{Katsuda2007}
	Katsuda, S., \& Tsunemi, H. 2007, Adv. Space Res. in press
\bibitem[Koyama et al.\ 2007]{Koyama2007}
	Koyama, K., et al. 2007, PASJ, 59S, 221
%\bibitem[Ku et al.\ 1984]{Ku1984}
%	Ku, W. H.-M., Kahn, S. M., Pisarki, R., \& Long, K. S. 1984,
%	ApJ, 278, 615			 
\bibitem[Levenson et al.\ 1997]{Levenson1997}
        Levenson, N. A. et al. 1997, ApJ, 484, 304
%\bibitem[Levenson et al.\ 1998]{Levenson1998}
%        Levenson, N. A., Graham, J. R., Keller, L. D., \& Richter,
%           M. J. 2002, ApJS, 118, 541
%\bibitem[Levenson et al.\ 1999]{Levenson1999}
%        Levenson, N. A., Graham, J. R., and Snowden, S. L. 1999, ApJ, 526, 874
\bibitem[Leahy 2004]{Leahy2004}
	Leahy, D. A. 2004, MNRAS, 351, 385
%\bibitem[Liedahl et al.\ 1995]{Liedahl1995}
%        Liedahl, D. A., Osterheld, A. L., \& Goldstein, W. H. 1995, 
%	ApJ, 438, L115 
%\bibitem[McCray \& Snow 1979]{McCray1979}
%        McCray, R. \& Snow, T. P., Jr. 1979, ARA\&A, 17, 213
\bibitem[Meyer et al.\ 1998]{Meyer1998}
	Meyer, D. M., Jura, M., \& Cardelli, J. A. 1998, ApJ, 493, 222
\bibitem[Miyata et al.\ 1994]{Miyata1994}
        Miyata, E., Tsunemi, H., Pisarski, R., \& Kissel, S. E. 1994,
          PASJ, 46, L101
%\bibitem[Miyata et al.\ 1998]{Miyata1998}
%        Miyata, E., Tsunemi, H., Kohmura, T., Suzuki, S., and Kumagai,
%        S. 1998, PASJ, 50, 257
\bibitem[Miyata et al.\ 1999]{Miyata1999}
        Miyata, E., \& Tsunemi, H. 1999, ApJ, 525, 305
\bibitem[Miyata et al.\ 2007]{Miyata2007}
        Miyata, E., Katsuda, S., Tsunemi, H., Hughes, J. P., Kokubun,
        M., \& Porter, F. S. 2007, PASJ, 59S, 163 (Paper {\scshape I})
\bibitem[Mitsuda et al.\ 2007]{Mitsuda2007}
	Mitsuda, K., et al. 2007, PASJ, 59, S1
\bibitem[Morrison \& McCammon 1983]{Morrison1983}
        Morrison, R., \& McCammon, D. 1983, ApJ, 270, 119
%\bibitem[Raymond et al.\ 2003]{Raymond2003}
%        Raymond, J. C., Ghavamian, P., Sankrit, R., Blair, W. P., and
%           Curiel, S. 2003, ApJ, 584, 770
\bibitem[Rauscher et al.\ 2002]{Rauscher2002}
        Rauscher, T., Heger, A., Hoffman, R. D., \& Woosley, S. E. 2002,
         ApJ, 576, 323
%\bibitem[Rudolph et al.\ 2006]{Rudolph2006}
%        Rudolph, A. L., Fich, M., Bell, G. R., Norsen, T., \& Simpson,
%	J. P. 2006, ApJS, 162, 346
%\bibitem[Serlemitsos et al.\ 2007]{Serlemitsos2007}
%        Serlemitsos, P. J., et al. 2007, PASJ, 59, S9
%\bibitem[Tsunemi et al.\ 1999]{Tsunemi1999}
%        Tsunemi, H., Miyata, E., \& Aschenbach, B. 1999, PASJ, 51, 711	
\bibitem[Tsunemi et al.\ 2007]{Tsunemi2007}
	Tsunemi, H., Katsuda, S., Nemes, N., \& Miller E. D., ApJ in press
%\bibitem[]{}
%
%\bibitem[]{}
%
\end{thebibliography}
\end{document}